\documentclass[%
 reprint,
superscriptaddress,
 amsmath,amssymb,
 aps,
]{revtex4-2}
\usepackage{comment}
\pdfoutput=1
\usepackage{xcolor}
\usepackage{graphicx}
\usepackage{dcolumn}
\usepackage{bm}
\usepackage[normalem]{ulem}
\usepackage{hyperref}

\begin{document}

\preprint{APS/123-QED}

\title{Existence of Corner Modes in Elastic Twisted Kagome Lattices}

\author{Hrishikesh Danawe}
\affiliation{Department of Mechanical Engineering, University of Michigan, Ann-Arbor, MI 48109-2125, United States}
\author{Heqiu Li}
\affiliation{Department of Physics, University of Michigan, Ann-Arbor, MI 48109-2125, United States}
\author{Hasan Al Ba'ba'a}%
\affiliation{Department of Mechanical Engineering, University of Michigan, Ann-Arbor, MI 48109-2125, United States}
\author{Serife Tol}
\affiliation{Department of Mechanical Engineering, University of Michigan, Ann-Arbor, MI 48109-2125, United States}

\date{\today}

\begin{abstract}
This Letter investigates the emergence of corner modes in elastic twisted Kagome lattices at a critical twist angle, known as self-dual point.  We show that a special type of corner modes exist and they are localized at a very specific type of corners independent of the overall shape of the finite lattice. Moreover, these modes appear even in the perturbed lattice at the corners with mirror planes. By exploring its counterpart in electronic system, we attribute such corner modes to charge accumulation at the boundary, which is confirmed by the plot of charge distribution in a finite lattice.
\end{abstract}

\maketitle

In the early 1990s, by drawing analogy to the quantum theory of solids, it is theoretically and experimentally shown that acoustoelastic waves propagating through a periodic array of scatterers exhibit frequency regimes in which the waves cannot propagate \cite{SIGALAS1993,Kushwaha1993,Sala1995}, analogous to the electronic bandgaps in semiconductors \cite{Leung1990,Zhang1990,Ho1990}. Since then, the study of periodically engineered artificial materials (a.k.a. phononic crystals and metamaterials) has matured into an exciting research field. The bandgap and dispersive properties of phononic crystals and metamaterials \cite{Page1996,Deymier2013,Laude2015} have resulted in unconventional approaches in various applications including sound and vibration attenuation \cite{Goffaux2003,Lin2021SMS}, cloaking \cite{Farhat2009}, lensing \cite{Jin2019,Danawe2020}, subwavelength imaging \cite{Sukovich2008}, and energy harvesting \cite{Tol2019Add}.

In recent years, the discovery of topological insulators in electronic systems has led the study of topological phases of phononic crystals inspired by their counterparts in photonic systems \cite{ma2019topological}. A mechanical system is said to be \textit{topological} or \textit{topologically protected} when certain conditions and symmetries are present in the dynamical system. The growing research interest in topological phases in mechanical systems stems from their special wave properties that offer immunity against defects and imperfections, no wave back-scattering, as well as unidirectional wave transmission \cite{ma2019topological,Zhang2018TopologicalSound}. Originally introduced in electronic/quantum systems, various types of topological states in mechanical systems have been studied. These include Quantum Hall Effect \cite{wang2015topological,chen2019mechanical,Nash2015TopologicalMetamaterials}, Quantum Spin Hall Effect \cite{Chaunsali2018ExperimentalResonators,Chen2018ElasticLattices,Chen2019TopologicalLattice}, Quantum Valley Hall Effect \cite{Chen2019TopologicalLattice,pal2017edge,AlBabaa2020Elastically-supportedInsulators,chen2018topological,liu2019experimental}, Majorana edge states \cite{Prodan2017DynamicalSystems}, edge states in polyatomic lattices \cite{AlBabaa2019DispersionCrystals}, and topological pumping in quasi-periodic systems \cite{Rosa2019EdgeLattices,Riva2020EdgePlates}. More recently, higher order topological insulators have been discovered, offering unique ability to exhibit in-gap zero-dimensional corner modes \cite{Serra-Garcia2018ObservationInsulator,Xue2019AcousticLattice,Fan2019ElasticStates}, which goes beyond regular one-dimensional edge states in other topological systems. 

\begin{figure}[]
\includegraphics{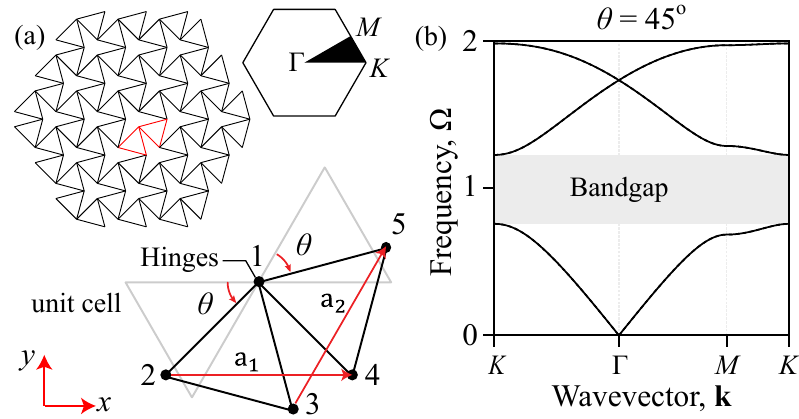}
\caption{\label{fig:fig1} (a) A finite structure of a self-dual Kagome lattice and its unit cell definition. The first Brillouin zone is depicted for reference. (b) Dispersion diagram of twisted Kagome lattice (i.e., $\theta=\pi/4$) showing three doubly degenerate dispersion branches. Note that $\Omega = \omega \sqrt{\frac{m}{k}}$.}
\end{figure}

This effort is particularly focused on corner modes emergence in a class of twisted Kagome lattices with in-plane motion. The considered twisted Kagome lattice is an example of Maxwell lattices, which have been at the heart of recent developments in the field of topological mechanics \cite{mao2018maxwell,zhang2018fracturing,Zhou2020SwitchableLattices,lubensky2015phonons}. The configuration of the Kagome lattice is shown in Fig.\ref{fig:fig1}, which is inspired from a recent study by Fruchart \textit{et al.} \cite{fruchart2020dualities}. A unit cell or  lattice is characterized by three discrete equal masses $m$, each of which moves in-plane in the $x$- and $y$- directions, making a total of six degrees of freedom per unit cell. The masses at the lattice sites are interconnected via central force bonds of stiffness $k$, and its shape is controlled by a twist angle $\theta$, with the perfect hinges at the nodes. As such, the unit cell dynamics exhibit a very interesting property, referred to as \textit{duality}, which means that two different configurations (or twist angles) can exhibit identical dispersion relations \cite{fruchart2020dualities}. The duality occurs about a critical angle $\theta_c = \pi/4$; hence, a lattice with twist angles $\theta_c\pm \delta \theta$ exhibit an identical dispersion relation. Such a critical point is referred to as a \textit{self-dual point}. The dispersion relation of a self-dual Kagome lattice ($\theta=\theta_c$) has three-distinct degenerate dispersion bands with two identical copies of dispersion branches on top of each other, computed by calculating the eigenfrequencies $\omega$ of the unit-cell's dynamical matrix at selected values of a wavevector $\mathbf{k} = \{ k_x,k_y \}$ (See Fig.\ref{fig:fig1} and Supplementary Materials (SM) for calculation details). The static and dynamic behaviour of elastic twisted Kagome lattices are well studied in literature. Hutchinson and Fleck \cite{HUTCHINSON2006756} investigated the collapse mechanisms (also known as floppy modes) of regular Kagome and strain-producing mechanisms in twisted Kagome lattice. Later, Sun \textit{et al.} \cite{Sun12369} studied the bulk and surface zero-frequency floppy modes of regular and twisted Kagome lattices under different boundary conditions. Recently, continuum approach is applied to explain the elastic behaviour of twisted Kagome lattice for zero frequency modes \cite{NASSAR2020104107} and topological edge soft modes in topological Kagome lattice \cite{sun2020}. More recently, Gonella \cite{Gonella2020} elucidated the dual nature of twisted Kagome lattice, originally put forward by Fruchart \textit{et al}.\cite{fruchart2020dualities}, using bidomain lattice structures and demonstrated the dependence of  duality behaviour on underlying mechanism of cell deformation. On the other hand, following the work of Xue \textit{et al.} \cite{Xue2019AcousticLattice} on acoustic higher order topological insulator on a Kagome lattice, Wu \textit{et al.} \cite{Wu2020} experimentally observed second order topologically protected corner states in regular elastic Kagome lattice. While rich in dynamics, the corner modes in elastic twisted Kagome lattices have not been previously studied in literature. It is, therefore, the aim of this study to demonstrate the existence of corner modes in twisted Kagome lattice and to explain their relationship to structural symmetries and bulk polarization, as well as to propound the necessary conditions for their manifestation. To this end, we also explore the analogous electronic system to explain the existence of corner modes via bulk polarization and Wannier centers.

\begin{figure}
\includegraphics{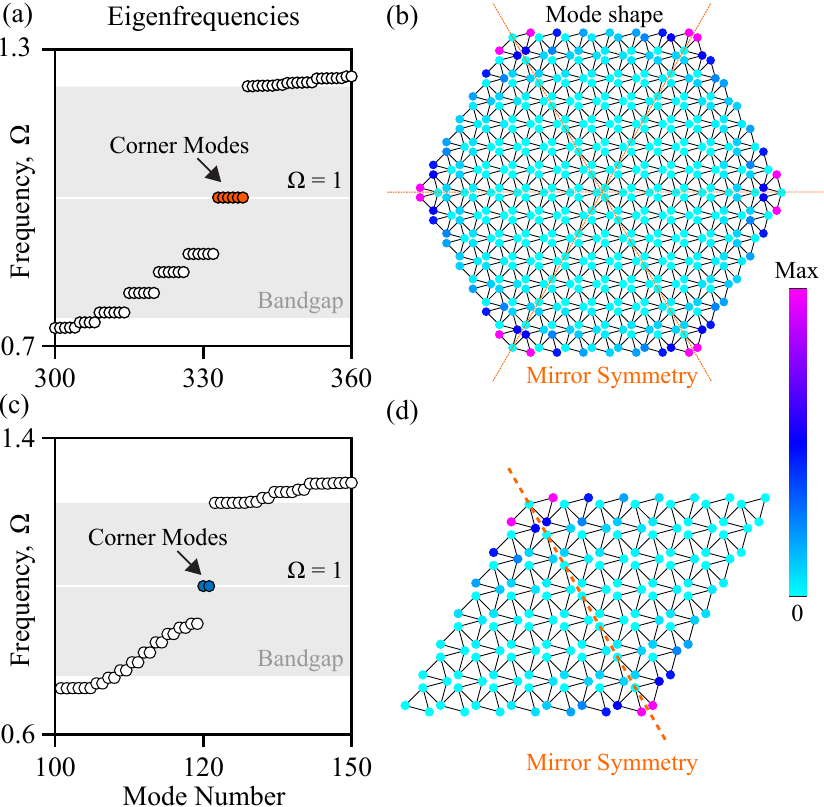}
\caption{\label{fig:fig2} (a) Eigenfrequency solutions for finite hexagon-shaped self-dual Kagome lattice with six corner modes at in-gap frequency of $\Omega=1$ and  (b) the mode shape for the corresponding in-gap corner modes localized at the corners of hexagon. (c) Eigenfrequency solutions for finite parallelogram-shaped self-dual Kagome lattice with two corner modes at in-gap frequency of $\Omega=1$ and  (d) the mode shape for the two in-gap corner modes localized at the two corners of parallelogram which lie on mirror symmetry plane.}
\end{figure}
\begin{figure*}
 \includegraphics[]{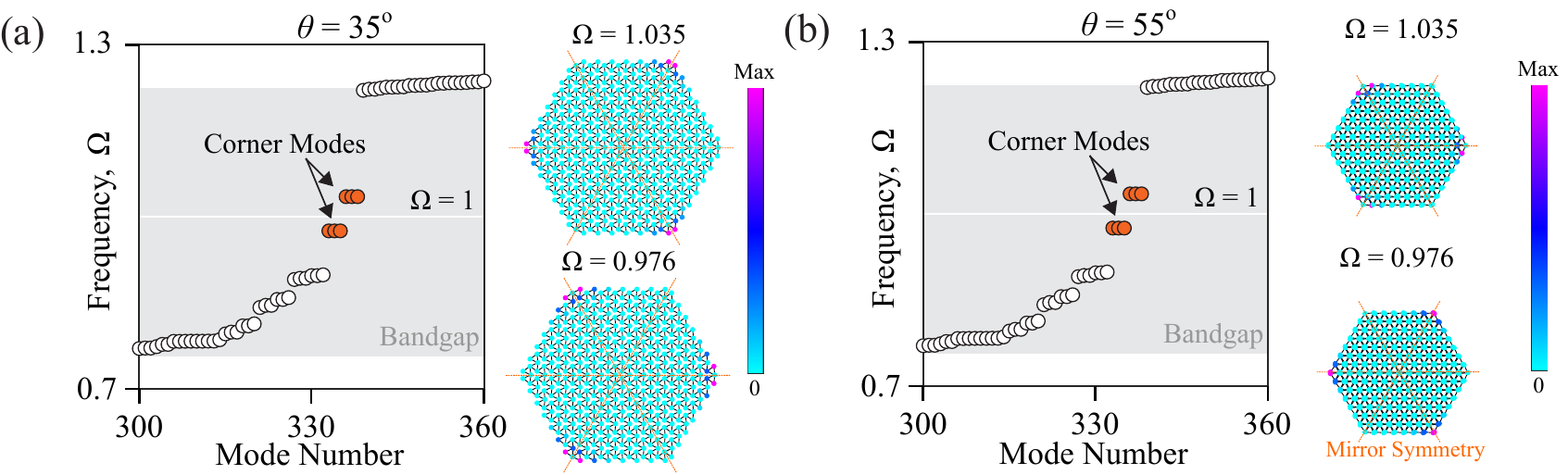}
  \caption{ Eigenfrequency solutions and modeshape of corner modes of finite hexagon-shaped twisted Kagome lattice for twist angles (a) $\theta=35^{\circ}$ and (b) $\theta=55^{\circ}$. Note that, the Kagome lattice shrinks with increasing twist angle as the two triangles in the unit cell come closer to each other; hence, the hexagonal lattice at $\theta=55^{\circ}$ is smaller than the hexagonal lattice at $\theta=35^{\circ}$.
 \label{fig:fig3}}
\end{figure*}
We first demonstrate the emergence of corner modes considering a finite self-dual Kagome lattice having a shape of regular hexagon as shown in Fig.~\ref{fig:fig1}(a). The self-dual Kagome lattice has three mirror symmetry planes oriented at angles $0$, $\pi/3$ and $2\pi/3$ (measured counterclockwise from a horizontal line), which pass through the corners of hexagonal structure. The degenerate dispersion band structure of the corresponding infinite lattice is shown in Fig.~\ref{fig:fig1}(b). We choose a hexagon lattice size of seven unit cells on each edge, which is sufficiently large to observe the in-gap corner eigenfrequncies, as shown in Fig.~\ref{fig:fig2}(a). There exist six corner modes (one for each corner) at in-gap frequency of $\Omega=\omega \sqrt{\frac{m}{k}}=1$, showing localization of deformation at the corners of the hexagon structure, a modeshape is depicted in Fig. \ref{fig:fig2}(b). The modeshape is obtained by linear combination of the six eigenmodes such that all corners are localized. Interestingly, the corresponding modeshapes of the corner modes show zero deformation of the lattice site lying on the mirror symmetry planes, a behavior that extends to all lattice sites of the same type in the vicinity of the corner. Other modes appearing in the bandgap (Fig. \ref{fig:fig2}(a)) are edge modes localized around the edges of hexagon, which appear in pairs due to the duality symmetry.

The appearance of corner modes at $\Omega = 1$, in fact, depends on the shape of the corner and the plane of symmetry. To demonstrate this dependence, consider a finite structure of an overall shape of a parallelogram, which only has a single mirror symmetry plane. While the structure has four total corners, only two corner modes appear at $\Omega = 1$ in the eigenfrequency calculations as depicted in Fig. \ref{fig:fig2}(c). As anticipated, the mode shapes of these corner modes are localized at the corners located at the symmetry planes as shown in Fig. \ref{fig:fig2}(d). This example emphasizes the importance of the existence of mirror symmetry for the manifestation of corner modes. An intriguing observation from both analyzed finite structure is that the corner modes always appear at a normalized frequency of $\Omega = 1$.  

The corner modes described above are specific to the self-dual point. The finite hexagonal and parallelogram shaped lattices away from self-dual point show corner modes appearing at different frequencies below and above the frequency $\Omega = 1$. For instance, the eigenfrequencies of corner modes for finite hexagonal along with the corner mode modeshapes for twist angles $\theta=35^{\circ}$ and $\theta=55^{\circ}$ are shown in Fig. \ref{fig:fig3}(a) and Fig. \ref{fig:fig3}(b), respectively. The bandgap shown is in between the 2nd and 3rd dispersion branch. The eigenfrequencies for the two configurations are exactly identical because of the duality and the corner modes appear in groups of three, one group at $\Omega=0.976$ and the other group at $\Omega=1.035$. The set of corners localized at the same frequency are of similar nature as can be seen from lattice geometry. Interestingly, the frequency at which the localization happens at a particular set of corners switch to other corner mode frequency while going from one configuration of the dual lattice pair to the other. For example, the localized set of corners at $\Omega=0.976$ for $\theta=35^{\circ}$ are similar to the localized set of corners at $\Omega=1.035$ for $\theta=55^{\circ}$. The transition happens about the self-dual point where all six corners are localized at a single frequency of $\Omega=1$. Furthermore, the deformation of lattice site of corner unit cells lying on mirror symmetry planes is not zero for the corner modes appearing at twist angles $\theta=35^{\circ}$ and $\theta=55^{\circ}$ which happens to be a characteristic feature at self-dual point only.

\begin{figure}
\includegraphics{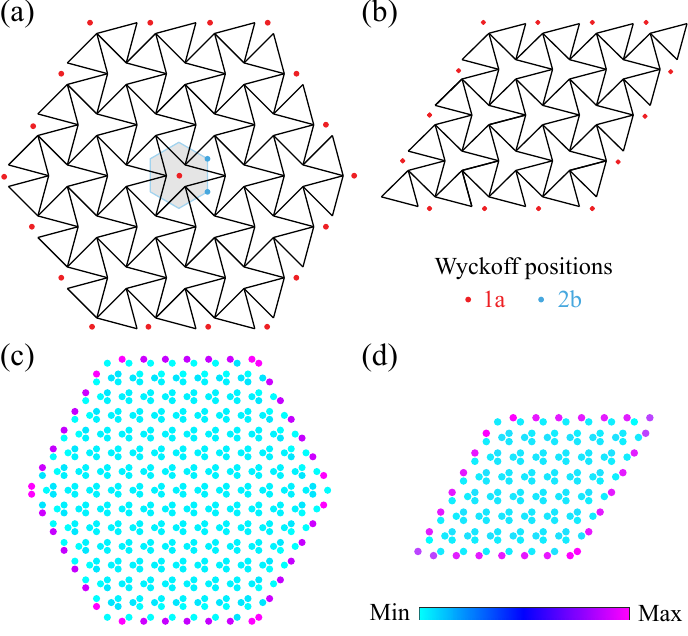}
\caption{\label{fig:fig4} (a) Hexagon-shaped lattice setup with all springs having the same spring constant. The unit cell is chosen to be the blue hexagon which preserves all crystalline symmetry. Wyckoff positions $1a$ and $2b$ are labeled as red and blue points, respectively. The exposed $1a$ positions at the boundary lead to edge and corner modes. (b) For the parallelogram lattice setup, the $60^\circ$ corner does not have exposed $1a$ positions, and the corner modes only appear at the $120^\circ$ corners, as shown in Fig.~\ref{fig:fig2}. Charge accumulation at edges and corners of (c) hexagon-shaped and (d) parallelogram-shaped lattice.}
\end{figure}

Next, we investigate the emergence of boundary and corner modes by drawing analogies to electronic topological insulators and analyzing the electron polarization and Wannier centers~\cite{Topchem,indicator,Benalcazar2017,Benalcazar2017b,Bernevig_HOTI,Benalcazar2019,Song2017}. Consider the hexagon-shaped lattice setup in Fig. \ref{fig:fig4}(a) in which all the bonds have the same spring constant. We can treat the dynamical matrix in momentum ($k-$) space as the Hamiltonian for some electronic systems. The system has crystalline symmetry $p31m$ (wallpaper group 15), in addition to time-reversal and duality symmetries. We choose the unit cell to be the blue hexagon in Fig.~\ref{fig:fig4}(a) which preserves threefold rotation $C_3$ and all the mirror symmetries. We next consider the electron polarization for the lowest two bands. $C_3$ symmetry quantizes the polarization to be $\mathbf P=ep(\mathbf a_1+\mathbf a_2)$ where $e$ is the electron charge, $\mathbf a_{1}$ and $\mathbf a_{2}$ are direct lattice primitive vectors (See Fig. \ref{fig:fig1}) and $p=0,1/3,2/3$ corresponds to Wyckoff positions $1a$ or $2b$ ~\cite{Benalcazar2019}. However, $2b$ position is not invariant under mirror symmetry, therefore, with both $C_3$ and mirror symmetry, the polarization must vanish. For the two lowest bands to have vanishing total polarization, their Wannier centers may either both locate at $1a$ position, or at the two distinct $2b$ positions. This can be further distinguished by analysing the symmetry representations of the two lowest bands at high symmetry momenta ($\Gamma$, ${K}$ and ${M}$ in Fig.~\ref{fig:fig1}) using the framework of topological quantum chemistry~\cite{Topchem}. 

The dynamical matrix $D(\mathbf k)$ in momentum space is computed in the SM. It has threefold rotational symmetry $C_3$ and three mirror symmetries. The high symmetry momenta $\Gamma$, $K$ and $-K$ are invariant under both $C_3$ and mirror symmetries, hence the little group there is $C_{3v}$. The high symmetry momentum $M$ is invariant only under one of the mirror symmetries, and the little group is $C_s$. By examining the eigenstates of $D(\mathbf k)$, we find that the lowest two bands at $\Gamma$, $K$ and $-K$ transform as the irreducible representation (irrep) $E$ of group $C_{3v}$, and at momentum $M$ the two bands have mirror eigenvalues $\pm 1$. This is identical to the elementary band representation generated by atomic states at Wyckoff position $1a$ that transform as irrep $E$ of the site symmetry group $C_{3v}$~\cite{Topchem}. Therefore, these bands have Wannier centers at $1a$. If the whole lattice has an integer number of unit cells, there will be no charge accumulation at the boundary because the Wannier centers are at $1a$ inside each unit cell. However, the finite lattice in Fig.~\ref{fig:fig4}(a) has fractional unit cells whose Wannier centers at $1a$ are exposed at the boundary, i.e., the red dots. These exposed Wannier centers lead to modes localized at the edges and corners. For the parallelogram setup in Fig. \ref{fig:fig4}(b), since there is no exposed $1a$ position at the $60^\circ$ corners, the corner modes only exist at the $120^\circ$ corners. 

Note that although the corner modes here are reminiscent of those in higher order topological insulators, the corner charge is not well-defined because of the non-zero edge charge. In this system, the bulk polarization measured in a symmetric unit cell (the hexagon in Fig. \ref{fig:fig4} (a)) is zero, and according to Ref.\cite{Benalcazar2019} it should not have edge polarization if the boundary contains complete unit cells. However, the boundary termination of the twisted Kagome lattice is incommensurate with the symmetric hexagon unit cell, i.e., there is no complete hexagon unit cells at the boundary, which is beyond Ref.\cite{Benalcazar2019}. Therefore, the edge and corner modes still emerge due to the incommensurate boundary termination although the bulk polarization vanishes. This is an example of termination-induced boundary modes.

The electronic charge distribution depicting the charge accumulation at the corners and sides of hexagon and parallelogram are as shown in Fig. \ref{fig:fig4}(c) and (d), respectively. The density is obtained at each lattice site by summing the norm of eigenvectors up to the corner modes at frequency $\Omega = 1$. The charge density is highest at the lattice sites of unit cells which form the $120^\circ$ angled corners. The charge accumulated on the edges of finite lattice is because of the edge modes in the bandgap.

\begin{figure}[b]
\includegraphics{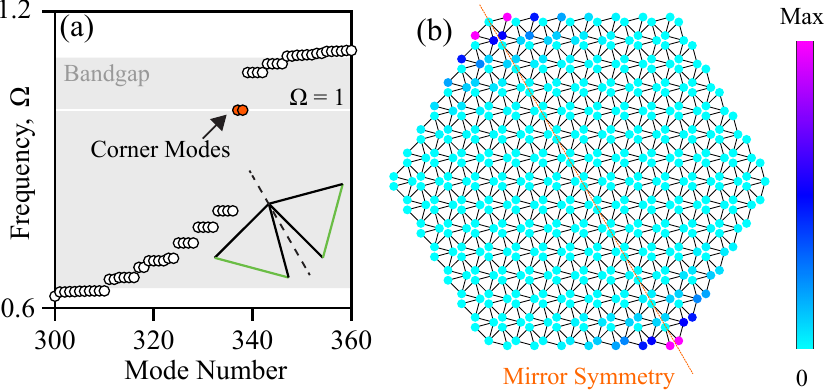}
\caption{\label{fig:fig5} (a) Eigenfrequenccies of perturbed hexagon-shaped lattice self-dual Kagome lattice, the stiffness of green links, as depicted in the inset, is made half of the original value (black links). (b) Modeshape of corner modes at $\Omega=1$, localization of the modes is at the corners which are on the mirror plane, the no. of mirror symmetry planes in the lattice reduces from 3 to 1 because of the perturbation.}
\end{figure}

Finally, we show an example of obtaining corner modes at specific corners by carefully perturbing the unit cell links. For the hexagonal structure analyzed in Fig.~\ref{fig:fig2}, we choose to reduce the stiffness of two links by a factor of $1/2$ (depicted by green lines in the inset of Fig. \ref{fig:fig5}(a)), such that a single mirror symmetry plane is retained. Interestingly, the dispersion relation of the perturbed lattice preserve degeneracy of the bands as well as duality, i.e., identical dispersion characteristics for lattices with twist angles equidistant from self-dual point. Note that, the frequency $\Omega=1$ is still in the first bandgap of perturbed lattice. Constructing a similar hexagonal finite lattice as in Fig.~\ref{fig:fig2}, yet with the perturbed links, the eigenfrequency solutions for which are shown in Fig. \ref{fig:fig5}(a). In this case, only two corner modes are observed at $\Omega=1$ and they are localized around the corners of hexagon, which lie on the sole mirror symmetry plane of the perturbed lattice. Once again, the mode shape shown in Fig.~\ref{fig:fig5}(b) is characterized by zero deformation of the lattice site in corner unit cells, which lie on mirror symmetry plane.

The corner modes observed in twisted Kagome lattices are reminiscent with the second-order topologically protected states in acoustic \cite{Xue2019AcousticLattice} and elastic \cite{Wu2020} regular Kagome lattices which rely on non-trivial bulk topology characterized by quantized Wannier centres. The corner modes observed in these studies are also shape dependent and appear only at corners with specific angles formed by the edges which cut through the Wannier centers. Despite these similarities, the origin of edge and corner modes in twisted Kagome lattices is distinct from the other systems. In twisted Kagome lattices, $C_3$ and mirror symmetries quantize Wannier centers to Wyckoff position 1a, leading to a vanishing bulk polarization. The edge and corner modes emerge due to the incompability between the lattice termination and bulk unit cell, even though the bulk of the system is a trivial atomic limit with vanishing polarization. The corner modes are found to be robust to any defects in the bulk (See SM for more details).

In summary, we demonstrated the emergence of corner modes in a self-dual Kagome lattice that appear at certain corners. Such corner modes were explained in light of an analogy to electronic insulators, which shows these boundary modes are induced by the lattice termination that is incompatible with bulk unit cell. Robustness of the self-dual Kagome lattices and their reconfigurable wave localization at specific corners may find novel approaches in various applications such as sensing and energy harvesting.

\begin{acknowledgments}
The authors extend their thanks to Dr.~Kai Sun for fruitful discussions on corner modes and Maxwell lattices.
\end{acknowledgments}

\bibliography{Bibliography.bib}

\end{document}


\title{Supplemental materials to the manuscript}
\maketitle

\section{Analytical Formulation and Dispersion Curve Calculations}
Figure 1 in the manuscript shows unit cell of twisted Kagome lattice with 5 nodes and direct lattice primitive vectors $\mathbf a_{1}$ and $\mathbf a_{2}$. Each node of the unit cell has two degrees of freedom $u$ and $v$ corresponding to the displacements in $x$- and $y$-directions. The unit cell has two triangles and the twist $\theta$ is acted in the clockwise (counterclockwise) direction for the upper (lower) triangle. Given the nature of elastic bonds, an equivalent stiffness matrix in the local coordinates (for deformation along the the bond) is given by the following form:

\begin{equation}
    \mathbf{K}_e = k
    \begin{bmatrix}
    1 & -1 \\ -1 & 1
    \end{bmatrix}
\end{equation}
where $k$ is the stiffness of each bond. Now, we define a transformation matrix which will facilitate the derivation of the stiffness matrix of the $i^{\text{th}}$ bond in $x-y$ coordinate system:
\begin{equation}
    \mathbf{\Phi}_i = 
    \begin{bmatrix}
    \cos(\phi_i)& \sin(\phi_i)& 0 & 0 \\
   0 & 0 & \cos(\phi_i)& \sin(\phi_i) 
    \end{bmatrix}
\end{equation}
where $\phi_i$ is the orientation of the $i^{\text{th}}$ bond with respect to $x$-axis and can be easily deduced from unit cell geometry. The angle $\phi_i$ takes one of the following values $\phi_i = \pm \theta, \frac{1}{3} \pi \pm \theta, \text{and} \frac{2}{3} \pi \pm \theta$. Such a transformation will define the coupling between the in-plane degrees of freedom $u_i$ and $v_i$ for the motion in the $x$ and $y$ directions, respectively. Making use of $\mathbf{\Phi_{i}}$, the stiffness matrix for the tilted bond is computed as:

\begin{equation}
    \mathbf{K}_i =  \mathbf{\Phi}_i^\text{T} \mathbf{{K}}_e  \mathbf{\Phi}_i = k
    \begin{bmatrix}
    \mathbf{A}_i & - \mathbf{A}_i \\
    -\mathbf{A}_i & \mathbf{A}_i
    \end{bmatrix}
\end{equation}
where

\begin{equation}
    \mathbf{A}_i = 
    \begin{bmatrix}
    \cos^2(\phi_i)& \frac{1}{2} \sin(2\phi_i) \\
   \frac{1}{2} \sin(2\phi_i) & \sin^2(\phi_i)
    \end{bmatrix}
\end{equation}

Upon assembling the global stiffness matrix, $\mathbf K_{c}$ and global mass matrix, $\mathbf M_{c}$ of the unit cell, we obtain at the following equation of motion:
\begin{equation}
    \mathbf{M}_c \mathbf{\ddot{z}} + \mathbf{K}_c \mathbf{z} =0
\end{equation}
where 
\begin{equation}
    \mathbf{z}^{\text{T}} = 
    \begin{Bmatrix}
    \mathbf{u}_1 & \mathbf{u}_2 & \mathbf{u}_3 & \mathbf{u}_4 & \mathbf{u}_5
    \end{Bmatrix}
\end{equation}
and $\mathbf{u}^{\text{T}}_\ell = \begin{Bmatrix}
u_\ell & v_\ell
\end{Bmatrix}$ is the degrees of freedom of the $\ell^{\text{th}}$ node. By virtue of the periodicity, the displacements of nodes 2-4 and 3-5 are related and governed by Bloch's theorem, such that:\begin{subequations}
    \begin{equation}
    \mathbf{u}_4 = \text{e}^{\textbf{i} \mathbf{k} \cdot \mathbf{a}_1}\mathbf{u}_2=\text{e}^{\textbf{i}k_1}\mathbf{u}_2
\end{equation}
\begin{equation}
    \mathbf{u}_5 = \text{e}^{\textbf{i} \mathbf{k} \cdot \mathbf{a}_2}\mathbf{u}_3=\text{e}^{\textbf{i}k_2}\mathbf{u}_3
\end{equation}
\label{eq:Bloch_conds}
\end{subequations}

\noindent
where $\mathbf k= (k_x, k_y)$ is the Bloch wave vector, $\mathbf a_{1}$, $\mathbf a_{2}$ are direct lattice primitive vectors and $k_{1}$, $k_{2}$ are reduced wave vectors given by $k_{1}=\mathbf k \cdot \mathbf a_{1}$, $k_{2}=\mathbf k \cdot \mathbf a_{2}$. As a result, one can apply a transformation $\mathbf{Q}$ and reduce the equations of motion to their essential degrees of freedom $\mathbf{z}_c^{\text{T}} = \begin{Bmatrix}
\mathbf{u}_1 & \mathbf{u}_2 & \mathbf{u}_3 
\end{Bmatrix}$, which is related to the original degrees of freedom via $\mathbf{z} = \mathbf{Q} \mathbf{z}_c$, where
\begin{equation}
    \mathbf{Q} = 
    \begin{bmatrix}
    \mathbf{I} & \mathbf{0} & \textbf{0} \\
    \mathbf{0} & \mathbf{I} & \textbf{0} \\
    \mathbf{0} & \mathbf{0} & \textbf{I}  \\
    \mathbf{0} & \mathbf{I}\text{e}^{\textbf{i}k_1} & \textbf{0} \\
    \mathbf{0} & \mathbf{0} & \textbf{I}\text{e}^{\textbf{i}k_2}\\
    \end{bmatrix}
\end{equation} 
The reduction of the equations of motion is achieved by substituting the relationship above and pre-multiplying by $\mathbf{Q}^{\text{H}}$, yielding the condensed unit cell equations of motion \cite{Hussein2014DynamicsOutlook}:
\begin{equation}
    \mathbf{\hat{M}}_c \mathbf{\ddot{z}}_c + \mathbf{\hat{K}}_c \mathbf{z}_c =0; \hspace{0.5 cm} \mathbf{\hat{M}}_c = \mathbf{Q}^{\text{H}}\mathbf{M}_c \mathbf{Q}, \hspace{0.5cm} \mathbf{\hat{K}}_c = \mathbf{Q}^{\text{H}}\mathbf{K}_c \mathbf{Q}
        \label{eq:EOM}
\end{equation}
where superscript H denotes the Hermitian transpose. Generally, such reduction of degrees of freedom renders the mass and stiffness matrices functions of the wave vector. Here, the mass matrix is a diagonal matrix and can be simply expressed as $\mathbf{\hat{M}}_c = m \mathbf{I}$, which is not a function of the non-dimensional wavenumbers. On the other hand, the stiffness matrix is a function of the non-dimensional wavenumbers and can be generally expressed as:
\begin{equation}
    \mathbf{\hat{K}}_c = k 
    \begin{bmatrix}
   \mathbf{\hat{K}}_{11} & \mathbf{\hat{K}}_{12} & \mathbf{\hat{K}}_{13} \\
    \mathbf{\hat{K}}_{12}^{\text{H}}& \mathbf{\hat{K}}_{22} & \mathbf{\hat{K}}_{23} \\
    \mathbf{\hat{K}}_{13}^{\text{H}} & \mathbf{\hat{K}}_{23}^{\text{H}} & \mathbf{\hat{K}}_{33} \\
    \end{bmatrix}
    \label{eq:K_matrix}
\end{equation}
The explicit form of the components in Eq.~(\ref{eq:K_matrix}) are:
\begin{subequations}
    \begin{equation}
    \mathbf{\hat{K}}_{11} = 
    \begin{bmatrix}
       \cos^2(\theta) + \frac{3}{2} & \sqrt{3}\left(\cos^2(\theta) - \frac{1}{2} \right) \\
       \sqrt{3}\left(\cos^2(\theta) - \frac{1}{2} \right) & \sin^2(\theta) + \frac{3}{2}
    \end{bmatrix}
\end{equation}
\begin{equation}
    \mathbf{\hat{K}}_{22} = 
    \begin{bmatrix}
       \cos^2(\theta) + \frac{3}{2} & \sqrt{3}\left(\frac{1}{2} - \cos^2(\theta) \right) \\
       \sqrt{3}\left(\frac{1}{2} - \cos^2(\theta) \right)  & \sin^2(\theta) + \frac{3}{2}
    \end{bmatrix}
\end{equation}
\begin{equation}
    \mathbf{\hat{K}}_{33} = 
    \begin{bmatrix}
      2 \sin^2(\theta) + 1 & 0 \\
       0  & 2 \cos^2(\theta) + 1
    \end{bmatrix}
\end{equation}
\begin{equation}
    \mathbf{\hat{K}}_{12} = -
    \begin{bmatrix}
       (1+e^{\textbf{i}k_1}) \cos^2(\theta) &
       \frac{\sin(2\theta)}{2} (1-e^{\textbf{i}k_1}) \\
       \frac{\sin(2\theta)}{2} (1-e^{\textbf{i}k_1})  & (1+e^{\textbf{i}k_1}) \sin^2(\theta)
    \end{bmatrix}
\end{equation}
\begin{equation}
    \mathbf{\hat{K}}_{13} = -
    \begin{bmatrix}
        \cos^2(\theta + \frac{\pi}{3}) + e^{\textbf{i} k_2} \cos^2(\theta - \frac{\pi}{3}) & -\frac{1}{4} \left(\sin(2\theta) \left(1-e^{\textbf{i}k_2} \right) - \sqrt{3} \cos(2\theta) \left(1+e^{\textbf{i}q_2} \right) \right) \\
        -\frac{1}{4} \left(\sin(2\theta) \left(1-e^{\textbf{i}k_2} \right) - \sqrt{3} \cos(2\theta) \left(1+e^{\textbf{i}k_2} \right) \right) & \sin^2(\theta + \frac{\pi}{3}) + e^{\textbf{i} k_2} \sin^2(\theta - \frac{\pi}{3})
    \end{bmatrix}
\end{equation}
\begin{equation}
    \mathbf{\hat{K}}_{23} = 
    \begin{bmatrix}
      \cos^2(\theta + \frac{\pi}{6}) - 1 - \cos^2(\theta + \frac{\pi}{3})e^{\textbf{i}(k_2-k_1)} & \frac{1}{4} \left(\sin(2\theta) \left(1-e^{\textbf{i}(k_2-k_1)} \right) + \sqrt{3} \cos(2\theta) \left(1+e^{\textbf{i}(k_2-k_1)} \right) \right) \\
      \frac{1}{4} \left(\sin(2\theta) \left(1-e^{\textbf{i}(k_2-k_1)} \right) + \sqrt{3} \cos(2\theta) \left(1+e^{\textbf{i}(k_2-k_1)} \right) \right) &  (\cos^2(\theta + \frac{\pi}{3}) - 1)e^{\textbf{i}(k_2-k_1)} - \cos(\theta + \frac{\pi}{6})^2
    \end{bmatrix}
\end{equation}
\end{subequations}
At self-dual point i.e. $\theta = 45^{\circ}$, the stiffness matrix simplifies to:
\begin{subequations}
    \begin{equation}
    \mathbf{\hat{K}}_{11} = \mathbf{\hat{K}}_{22} = \mathbf{\hat{K}}_{33} = 2 \mathbf{I}
\end{equation}
\begin{equation}
    \mathbf{\hat{K}}_{12} = - \frac{1}{2}
    \begin{bmatrix}
       (1+e^{\textbf{i}k_1}) &
        (1-e^{\textbf{i}k_1}) \\
       (1-e^{\textbf{i}k_1})  & (1+e^{\textbf{i}k_1})
    \end{bmatrix}
\end{equation}
\begin{equation}
    \mathbf{\hat{K}}_{13} =
    \begin{bmatrix}
        \frac{\sqrt{3}}{4}\left(1-e^{\textbf{i}k_2} \right) - \frac{1}{2}\left(1+e^{\textbf{i}k_2} \right) & \frac{1}{4} \left(1-e^{\textbf{i}k_2} \right) \\ \frac{1}{4} \left(1-e^{\textbf{i}k_2} \right) & -\frac{\sqrt{3}}{4}\left(1-e^{\textbf{i}k_2} \right) - \frac{1}{2}\left(1+e^{\textbf{i}k_2} \right)
    \end{bmatrix}
\end{equation}
\begin{equation}
    \mathbf{\hat{K}}_{23} = 
    \begin{bmatrix}
      -\frac{\sqrt{3}}{4}\left(1-e^{\textbf{i}(k_2-k_1)} \right) - \frac{1}{2}\left(1+e^{\textbf{i}(k_2-k_1)} \right) & \frac{1}{4} \left(1-e^{\textbf{i}(k_2-k_1)} \right) \\
      \frac{1}{4} \left(1-e^{\textbf{i}(k_2-k_1)} \right) &  \frac{\sqrt{3}}{4}\left(1-e^{\textbf{i}(k_2-k_1)} \right) - \frac{1}{2}\left(1+e^{\textbf{i}(k_2-k_1)} \right)
    \end{bmatrix}
\end{equation}
\label{eq:stiffMat_criticalpoint}
\end{subequations}
By substituting harmonic solution $\mathbf z_c=\mathbf z_{c0} e^{\textbf{i}\omega t}$ in Eq. \ref{eq:EOM}, we obtained an eigenvalue problem as follows:
\begin{equation}
    \mathbf{-\hat{M}}_c \omega^2 + \mathbf{\hat{K}}_c  =\mathbf 0;
    \label{eq:EIG}
\end{equation}

The dynamical matrix is given by:
\begin{equation}
   D_0(\mathbf k)= \mathbf{\hat{M}}_c^{-1}  \mathbf{\hat{K}}_c  =\frac{1}{m}  \mathbf{\hat{K}}_c ;
    \label{eq:DynamicMatrix}
\end{equation}

\begin{figure}
 \includegraphics[width=1.5 in]{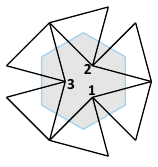}
  \caption{ The hexagon represents a unit cell that is symmetric under $C_3$ and mirror symmetries. This symmetric unit cell is used in the bulk polarization computation.}
 \label{fig:hexa}
\end{figure}

To compute the bulk polarization, we need to choose the hexagonal unit cell that is symmetric under $C_3$ and mirror symmetries as in Fig.~\ref{fig:hexa}. This change of unit cell corresponds to a transformation $D(\mathbf k)=U(\mathbf k)D_0(\mathbf k)U(\mathbf k)^{\dagger}$, where $U(\mathbf k)=diag\{1,1, e^{i k_2}, e^{i k_2},e^{i (k_2-k_1)}, e^{i (k_2-k_1)} \}$, where $diag$ means diagonal matrix. Then the dynamical matrix $D(\mathbf k)$ used for bulk polarization calculation in the manuscript is:

\begin{widetext}
\begin{eqnarray}
&&D(\mathbf k)=\frac{k}{m}
\left( \begin{array}{cccccc}
    2 & 0 & -\frac{z_1}{2} & -\frac{z_2}{2} & \frac{\sqrt{3}}{4}z_4-\frac{1}{2}z_3 & \frac{1}{4}z_4 \\
    0 & 2 &  -\frac{z_2}{2} & -\frac{z_1}{2} & \frac{1}{4}z_4 & -\frac{\sqrt{3}}{4}z_4-\frac{1}{2}z_3 \\
    -\frac{\overline{z}_1}{2} & -\frac{\overline{z}_2}{2} & 2 & 0 & -\frac{\sqrt{3}}{4}z_6-\frac{1}{2}z_5 & \frac{1}{4}z_6 \\
    -\frac{\overline{z}_2}{2} & -\frac{\overline{z}_1}{2} & 0 & 2 & \frac{1}{4}z_6 & \frac{\sqrt{3}}{4}z_6-\frac{1}{2}z_5  \\
    \frac{\sqrt{3}}{4}\overline{z}_4-\frac{1}{2}\overline{z}_3 & \frac{1}{4}\overline{z}_4 & -\frac{\sqrt{3}}{4}\overline{z}_6-\frac{1}{2}\overline{z}_5 & \frac{1}{4}\overline{z}_6 & 2 & 0 \\
    \frac{1}{4}\overline{z}_4 & -\frac{\sqrt{3}}{4}\overline{z}_4-\frac{1}{2}\overline{z}_3 & \frac{1}{4}\overline{z}_6 & \frac{\sqrt{3}}{4}\overline{z}_6-\frac{1}{2}\overline{z}_5 & 0 & 2
  \end{array}
\right) \\
&&z_1=e^{-ik_2}+e^{i(k_1-k_2)},\  z_3=e^{i(k_1-k_2)}+e^{ik_1},\ z_5=e^{ik_1}+e^{ik_2},\nonumber\\ 
&&z_2=e^{-ik_2}-e^{i(k_1-k_2)},\ z_4=e^{i(k_1-k_2)}-e^{ik_1},\  z_6=e^{ik_1}-e^{ik_2},\ k_1=\mathbf{k}\cdot \mathbf{a}_1,\ k_2=\mathbf{k}\cdot \mathbf{a}_2 \nonumber
\end{eqnarray}
\end{widetext}

The basis is chosen to be $(u_{1x},u_{1y},u_{2x},u_{2y},u_{3x},u_{3y})$ where $u$ represents the displacement of the three sites in each unit cell as labeled in Fig.~\ref{fig:hexa}. The system has threefold rotational symmetry $C_3$ and three mirror symmetries. Denote the horizontal mirror as $M_y$. The action of these symmetries on momentum $\mathbf k=(k_x,k_y)$ is given by $C_3\mathbf k=(-\frac{1}{2}k_x-\frac{\sqrt{3}}{2}k_y,\frac{\sqrt{3}}{2}k_x-\frac{1}{2}k_y)$ and $M_y\mathbf k=(k_x,-k_y)$. The matrices representing these symmetry operators are
\begin{eqnarray}
C_3&=&
\left( \begin{array}{cccccc}
    0 & 0 & 0 & 0 & -\frac{1}{2} & -\frac{\sqrt{3}}{2} \\
    0 & 0 & 0 & 0 & \frac{\sqrt{3}}{2} & -\frac{1}{2}\\
    -\frac{1}{2} & -\frac{\sqrt{3}}{2} & 0 & 0 & 0& 0\\
    \frac{\sqrt{3}}{2} & -\frac{1}{2} & 0 & 0 & 0 & 0\\
    0 & 0 & -\frac{1}{2} & -\frac{\sqrt{3}}{2} & 0 & 0\\
    0 & 0 & \frac{\sqrt{3}}{2} & -\frac{1}{2} & 0 & 0
\end{array}
\right) \\
M_y&=&
\left( \begin{array}{cccccc}
    0 & 0 & 1 & 0 & 0 & 0 \\
    0 & 0 & 0 & -1 & 0 & 0\\
    1 &0 & 0 & 0 & 0& 0\\
    0 & -1 & 0 & 0 & 0 & 0\\
    0 & 0 & 0 & 0 & 1 & 0\\
    0 & 0 & 0 & 0 & 0 & -1
\end{array}
\right)
\end{eqnarray}
The dynamical matrix satisfies the symmetry constrains $M_yD(\mathbf k)M_y^{-1}=D(M_y\mathbf k)$ and $C_3D(\mathbf k)(C_3)^{-1}=D(C_3\mathbf k)$.

\begin{figure}[h!]
  \includegraphics[]{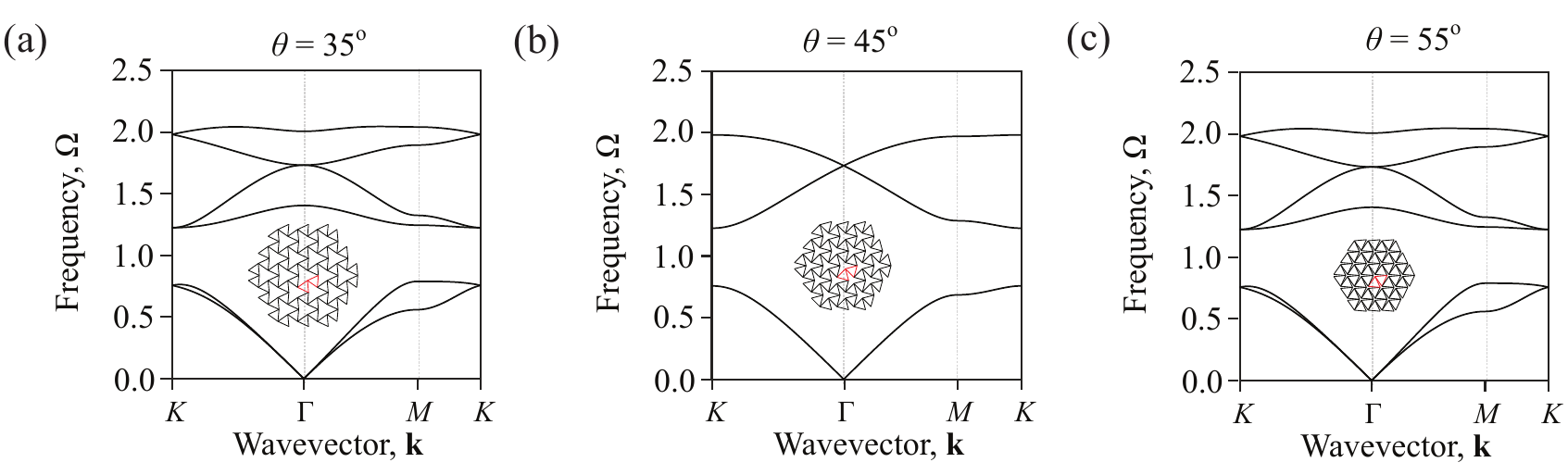}
  \caption{ The dispersion band structure of twisted Kagome lattice for (a) $\theta=35^{\circ}$ (b) $\theta=45^{\circ}$ (c) $\theta=55^{\circ}$. The lattice structures are shown in inset with unit cell highlighted in red.  Note that $\Omega = \omega \sqrt{\frac{m}{k}}$\label{fig:S1}}
\end{figure}

\begin{figure}[h!]
  \includegraphics[]{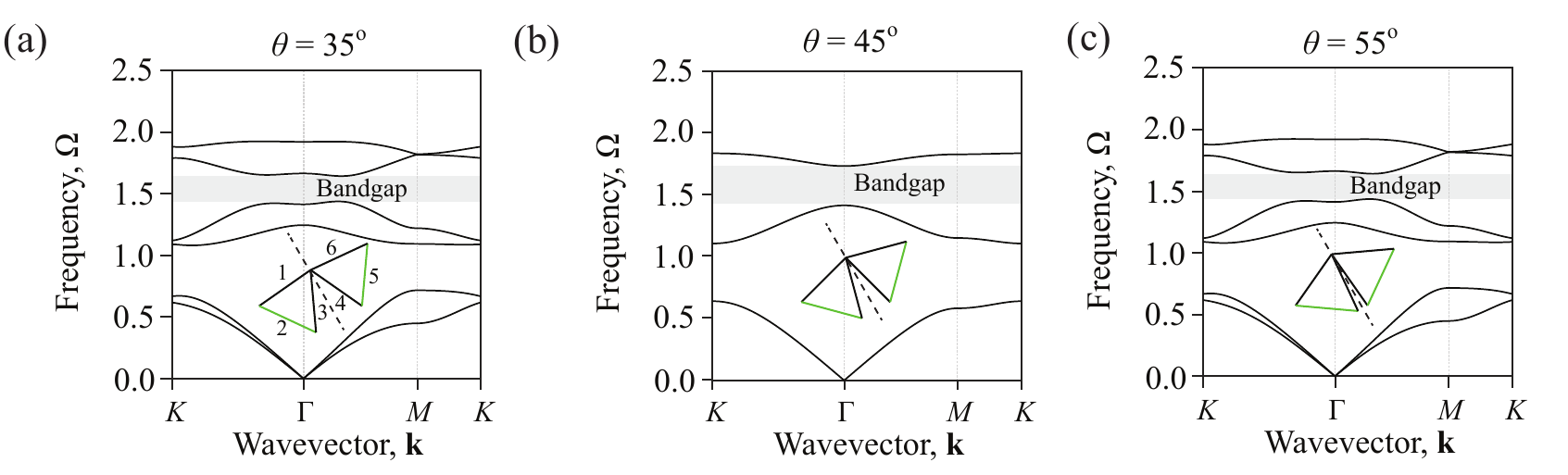}
  \caption{ The dispersion band structure of perturbed twisted Kagome lattice for (a) $\theta=35^{\circ}$ (b) $\theta=45^{\circ}$ (c) $\theta=55^{\circ}$. The unit cells along with the leftover mirror symmetry planes (dotted black lines) are shown in inset with perturbed bonds colored in green. The stiffness of green bonds is half that of the black bonds. A bandgap is highlighted which opens between 4th and 5th dispersion branch because of the perturbation. \label{fig:S2}}
\end{figure}

Solving Eq. \ref{eq:EIG} for specific $(k_1,k_2)$ values which lie on the boundary of irreducible Brillouin zone $K-\Gamma-M-K$ result is dispersion band structure as shown in Fig. \ref{fig:S1} for three different twist angles. The twist angles equidistant from the self dual point, i.e. $\theta = 35^{\circ}$  and $\theta = 55^{\circ}$, produce identical dispersion band structure even when the geometry of the lattices, as shown in the insets of Fig. \ref{fig:S1}, are completely different. Also, at self-dual point, only three branches are visible in the dispersion band structure which are in fact six branches with two identical copies of each branch lying on top of each other. Thus, this special twist angle of $\theta = 45^{\circ}$ is know as self-dual point \cite{fruchart2020dualities}. Furthermore, the duality and degeneracy of twisted Kagome lattice is still preserved even when all the bonds of a triangle are of different stiffness given that the bond pairs 1-4, 2-5 and 3-6 of the two triangles in the unit cell have identical stiffness. Additionally, a bandgap opens between the 4th and 5th dispersion branch when the bonds are perturbed. The dispersion band structure obtained for a special case when only the stiffness of bond pair 2-5 is half that of the other bond pairs is shown in Fig. \ref{fig:S2}. The duality about the self-dual point and the band degeneracy at the self-dual point are evident in the plots.

\section{Robustness of corner modes against defects }

\begin{figure}[ht]
  \includegraphics[]{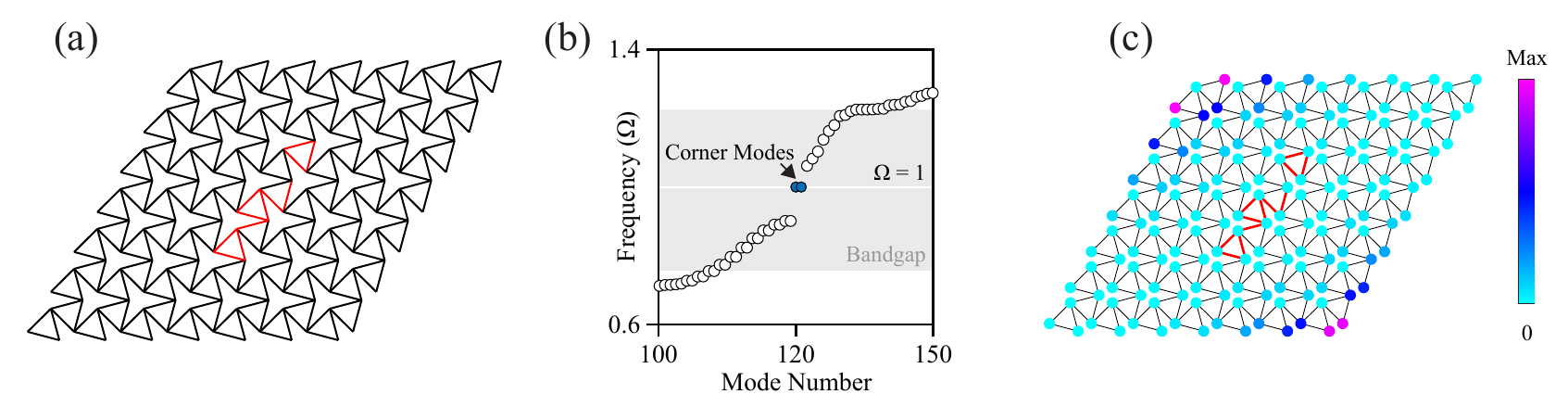}
  \caption{ (a) A finite parallelogram-shaped self-dual Kagome lattice with random defects, stiffness of red links is half that of black links. (b) Eigenfrequency solutions and (c) the modeshape for the two in-gap corner modes of the defective lattice. \label{fig:S4}}
\end{figure}

The corner modes observed in twisted Kagome lattices are robust against defects present in the bulk of finite lattice. In order to demonstrate the robustness, we introduced random defects in a parallelogram shaped finite self-dual Kagome lattice by reducing the link stiffness of the red links as shown in Fig. \ref{fig:S4}(a). The two corner modes still exist at $\Omega=1$ in the bandgap as shown in Fig. \ref{fig:S4}(b). Introduction of defects results in additional in-gap modes above $\Omega=1$ which are localized around the defects. However, the corner modes are not affected by the bulk defects and the 120$^\circ$ corners of the finite lattice are localized in the modeshape as depicted in Fig. \ref{fig:S4}(c). The modeshape is identical to the modeshape of lattice with no defects.

\bibliography{Bibliography.bib}

